# FOSS solution for Molecular Dynamics Simulation Automation and Collaboration with MDSGAT

*J. G. Nelson* [a,b], *X. Liu* [a,b,c] *and K. T. Yong*[*a,b,c]

a. School of Biomedical Engineering, The University of Sydney, Sydney, New South Wales 2006, Australia.

b. The Biophotonics and Mechanobioengineering Laboratory, The University of Sydney, Sydney, New South Wales 2006, Australia.

c. The University of Sydney Nano Institute, The University of Sydney, Sydney, New South Wales 2006, Australia.

**\*Corresponding Author:** ken.yong@sydney.edu.au

**Abbreviations:**

FTP – File Transfer Protocol

MD – Molecular Dynamics

MDS- Molecular Dynamics Simulation

PBS – Portable Batch System

**Figures:**

# Abstract:

The process of setting up and successfully running Molecular Dynamics Simulations (MDS) is outlined to be incredibly labour and computationally expensive with a very high barrier to entry for newcomers wishing to utilise the benefits and insights of MDS. Here, presented, is a unique Free and Open-Source Software (FOSS) solution that aims to not only reduce the barrier of entry for new Molecular Dynamics (MD) users, but also significantly reduce the setup time and hardware utilisation overhead for even highly experienced MD researchers. This is accomplished through the creation of the Molecular Dynamics Simulation Generator and Analysis Tool (MDSGAT) which currently serves as a viable alternative to other restrictive or privatised MDS Graphical solutions with a unique design that allows for seamless collaboration and distribution of exact MD simulation setups and initialisation parameters through a single setup file. This solution is designed from the start with a modular mindset allowing for additional software expansion to incorporate numerous extra MDS packages and analysis methods over time.

# 1. Introduction

## Background

Molecular Dynamics Simulation (MDS) is a computational biophysical modelling method, typically used to simulate the interactions between various macro biomolecules such as proteins, lipids and cellular membranes, however, more recent use has seen the use cases of MDS expanding towards additional fields and uses such as nanomaterials [1] and synthetic chemistry [2], [3]. The operating principle behind how MD Simulations function is by implementing Newtown's laws of motion on all atoms contained within a simulated system for each simulated time step [4]. For best results and accuracy smaller time steps are often used to mitigate error and approach closer to 'real-time' simulation with a common range for each time step typically falling anywhere between 1-5 femtoseconds, depending on simulation volatility. The actual implementation of these equations is called a forcefield and accounts for atom distances and bond lengths, angles and dihedral angles. For reference the OPLS forcefield equation [5] is shown below and depicts all the common aspects of a typical modern forcefield [6]:

$$E = \sum_{i<j} \left[ \frac{q_i q_j e^2}{r_{ij}} + 4\varepsilon_{ij} \left( \frac{\sigma_{ij}^{12}}{r_{ij}^{12}} - \frac{\sigma_{ij}^{6}}{r_{ij}^{6}} \right) \right] f_{ij} + \sum_{bonds} K_r (r - r_{eq})^2 + \sum_{angles} K_\theta (\theta - \theta_{eq})^2 +$$
$$\sum_{dihedrals} \left[ \frac{V_1}{2}(1 + \cos\varphi) + \frac{V_2}{2}(1 - \cos 2\varphi) + \frac{V_3}{2}(1 + \cos 3\varphi) + \frac{V_4}{2}(1 - \cos 4\varphi) \right]$$

In order to perform this calculation, a detailed dataset of atom characteristics must be supplied in order to have enough information to calculated atomic forces, for this purpose, many stand-alone "MDS Packages" have been created that both supply this information alongside typically also managing a number of practical operating tasks. Such tasks include hardware allocations and optimisations as well as data logging and file packaging into workable formats for user input and output. Many of these MDS Packages exist with some common options including GROMACS [7] CHARMM [8], AMBER [9] and NAMD [10], however, AMBER and CHARMM are more specialised for their respective dedicated forcefields whereas GROMACS and NAMD are more universal packages allowing for a wide array of forcefields to be utilised. Most modern and commonly utilised forcefields such as AMBER and OPLS will operate with an acceptable level of accuracy for the majority of simulation tasks with each forcefield having slight biases in certain macromolecule characteristics and formations such as AMBERs and CHARMMs slight bias towards forming helical structures or the original GROMACS forcefield GROMOS bias towards beta sheets.

MDS is a rapidly advancing technique that is seeing more and more widespread use as hardware capabilities steadily increase and costs are lowering. Some notable uses of MDS in recent day include the modern process of drug design and pre-evaluation, the refinement of experimentally or machine learning derived protein structures, characteristic analysis of biomolecular flexibility and mobility under various conditions or protein interactions and allostery [4], [11], [12].

However, despite costs lowering and capabilities growing from where they began, MDS is still some distance away from being commonly utilised in many facilities, one factor is the high upfront cost with entry level MDS capable systems beings worth at least $5000 USD [13], [14], quickly rising to tens of thousands of $USD and still taking weeks to simulate a microsecond of data for large simulations but another crucial factor is usability and training. Despite the prolonged simulation time, it is still a solvable issue thanks to both the ability to acquire more performant hardware but also job parallelisation allowing for multiple simulations to be ran simultaneously. The usability and training issue however, often contributes a significant portion of the time required to run MD simulations with inexperienced users taking weeks to months to fully grasp how to create and setup even basic simulations with advanced users still struggling to create more complicated simulations such as those involving multiple solvents, molecules or specific initial starting conditions which can take days or sometimes even weeks to troubleshoot.

## Current Challenges with MD Simulations

This difficulty in MDS operation is typically only worsened by standard hardware allocation schemes that delay bug testing and simulation experimentation by shared access queue wait times which can range from hours to weeks depending on hardware availability and institutional development. These factors compound on one another making MDS a rather impractical research method for any researchers who are trying to get started as it can take months to produce any meaningful results ready for publication. The primary driver behind this complexity is the nature of MDS setup and construction, as this process involves the creation of several manual configuration and runtime files for simple simulations and often requiring some level of specific mid-simulation editing of these files for more advanced configurations. All without any pre-error check processing or template scaffolding for simulation setup. Additionally, support for complex analysis is often unsupported by MDS packages, often requiring users to have to program their own analysis tools using plotting and analysis libraries available in code languages such as Python or MATLAB which adds additional complexity and burden on the user resulting in further delays to findings from any simulations.

All of these error prone steps often lead to a non-insignificant number of failures when attempting to run an MD simulation at various stages throughout the simulation cycle meaning that not only is researcher and user time being utilised in the creation and troubleshooting of these simulations but also a significant level of hardware uptime is utilised on non-meaningful simulation data which can serve as a significant opportunity cost for bulk computing devices that service a large number of users as is commonplace in many academic institutions.

## Objectives

The proposed solution to these issues presented here is in the creation of a simplified graphical interface in which users can interact with simulation parameters more naturally and have their inputs pre-error checked prior to simulation. This avoids a number of issues including reducing the upfront difficulty and training required, reducing the debug and simulation test times, ensuring simulation results are sound and minimising wasted hardware utilisation. Whilst there are numerous benefits to implementing such an interface only a small handful of groups have attempted to do so with many of these implementations having significant limitations such as lacking any analysis support to being locked into a single global server for simulation processing or even only being limited to a single MDS package and forcefield combination.

This leads there to be no current solution that is attempting to confront the MDS difficulty problem as a whole and as such this paper is the first iteration at an attempt to create a unified MDS program that can efficiently and easily create, configure and analyse MD simulations for various conditions across various hardware configurations.

To realise and overcome this problem the program MDSGAT (Molecular Dynamics Script Generator and Analysis Tool) has been founded with several primary goals, first and foremost, create a simplified user GUI that can be used to automate a number of repetitive tasks or parameters required when configuring a simulation to achieve easy functionality and accuracy without removing the users control over specific parameters when needed, this will serve to drastically lower the barrier of entry into MDS. Secondly allow for local configuration and analysis of MD simulations from the users own machine, this is to minimise required contact time with centralised computing hardware as well as aid in the tracking and user management/storage of research data and materials. Third and last of our primary objectives is the creation of a uniform distribution file that contains MD simulation parameters and required data for easy distribution and initiation of MD simulations across different hardware configurations allowing for easier collaboration and cross validation during publication whilst also serving as a method of creating a datastore of valid simulation setups for other users to learn from.

# 2. Existing Work

## Current Tools for MD Simulations

Currently, the standard method for running MD simulations falls into one of two categories, either the user is required to learn the full complexity of operating an MDS package which involves the creation and troubleshooting of a large number of simulation configuration and run files whilst also managing optimal hardware allocations, system setup and command methodology through the command line interface. This approach is most inline with the base installation GROMACS, AMBER and NAMD packages which contain a number of programs to aid in the creation and runtime of a simulation through command line interface. Alternatively, there are a small number of programs available that provide a graphical interface to aid in the setup and analysis of MD simulations, however, each of these programs appears to be highly specialised in nature only designed to be functional with a single MDS package or forcefield [15], [16] or restricted to using the GUI providers hardware or computing server [17]. Additionally, only a very small number of currently available MD simulation assistance tools attempt to introduce a solution to the issue of reproducibility with some journals considering mandatory guidelines [18], [19] for the publication of MDS related papers and numerous researchers calling on standardisation for the release of simulation parameters and configurations alongside additional practices to be introduced to aid in reproducibility [4], [20].

## Impact of Graphical Solutions in Computational Science

Since the creation of computer programming and computation the problem of user accessibility and experience has been constant with the now commonplace and accepted settled user experience being primarily through the use of Graphical User Interfaces (GUIs). The use of a GUI allows for an inexperienced user of a program or toolset to more intuitively perform tasks often without requiring an in-depth understanding of the underlying functions at hand. Additionally, GUIs serve as a powerful tool for advanced users allowing for a rapid method of task navigation that is significantly more difficult to emulate using traditional command line interface methods of input. Through the implementation of graphical methods for various software suites including scientific computing the world has a seen a tremendous uptick is early adoption and training of various scientific principles and methods facilitated by powerful and intuitive GUIs [21]. Such examples include PyMOL [22], [23] as an incredibly versatile suite for molecular modelling or LTSpice [24], [25] for in-depth electrical circuit simulation and analysis, both of which allows for a level of conceptualisation and understanding that would be incredibly difficult to mimic without a visual feedback system to the user [26].

# 3. System Architecture

## MDS Overview

The creation and setup of an MD simulation contains numerous systems and files that each need to be manually modified prior to simulation start. The following will outline each of these files and stages required and is based on the GROMACS MDS package for both consistency and to be representative of the initial MDS package supported by MDSGAT on v1.0 release. A brief overview of the process for MDS using the GROMACS package is summarised in the figure below:

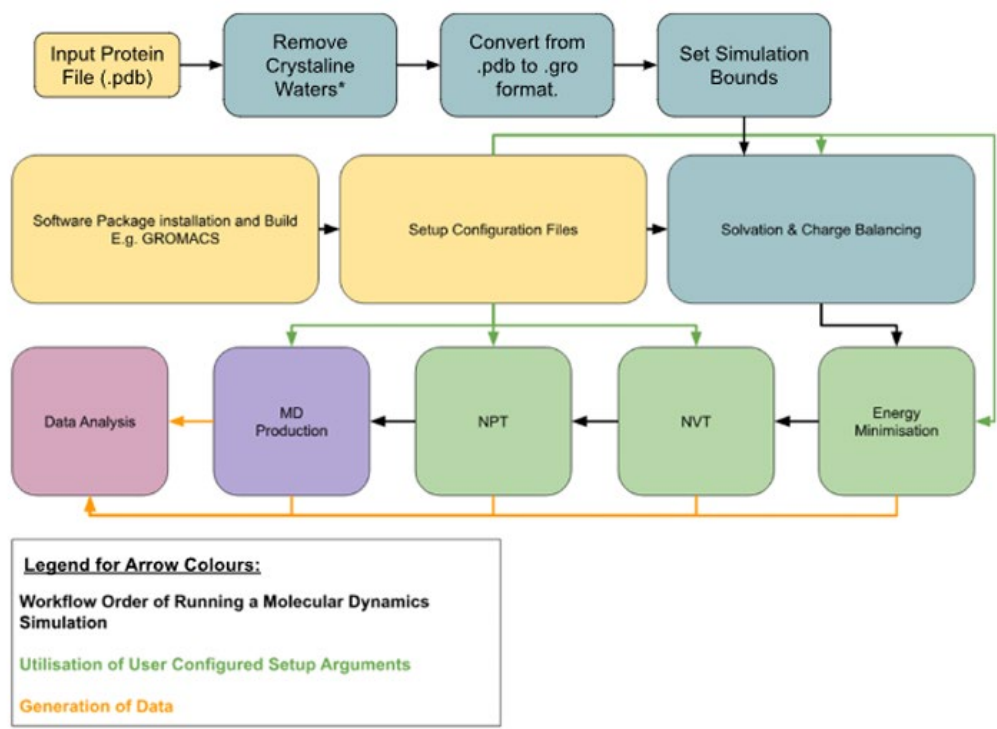

*Figure 1. Generalised overview of the MDS creation and run process using the GROMACS MDS package as an example.*

## Files:

**Protein/Molecule Structure file (.pdb, .gro, .g96, .brk, .ent, .esp, .tpr):**

This is typically an experimentally or machine learning derived molecule structure file that includes information such as atom positions and molecular structure. The most commonly used format is the protein databank format (.pdb) with files commonly found on databases such as RCSB PDB [27] or the Alphafold Protein Structure Database [28]. Some of these file formats such as gromacs (.gro) are also used to capture the system structure of a simulation and used to provide a "snapshot" of the atom positions at certain timesteps.

**Included topology files (.itp):**

Outlines atomic and molecular bond and force information to be used by the simulation software. Typically paired with both forcefield and MDS package suppliers to provide atomic force and solvent characteristics in-line with expected and tested characteristics of each package, for example, a forcefield expecting a 3-point water model would not function when supplied with a 4-point water .itp file.

**Configuration files (.mdp):**

Contains most of the initial and runtime parameters required for the simulation to function for a standard simulation a total of 5 configuration files would be required with varying number of parameters; ion configuration (~11 parameters), energy minimisation configuration (~11 parameters), NVT configuration (~31 parameters), NPT configuration (~35 parameters) and MD production run configuration (~36 parameters). Common parameters include number of timesteps, timestep size (duration), log write frequency, system constraints, temperature and pressure coupling algorithms and electrostatic modelling methods and ranges.

**Portable trajectory format (.xtc):**

Stores information relating to atom positions across the entirety of the simulations duration, allowing for visualisation of atom positions over time. Also stores some rudimentary information relating to atom forces and velocities for analysis purposes.

**Topology files (.top)**

Combined force and bond listing file that includes directories pointing to included topology files (.itp files) alongside direct force and bond information of newly updated or included structures such as molecule or solvent atoms. Includes data such as atom count, charge and order alongside bond information such as angle dihedral angle and atomic pairs.

**Project Allocation/Run Script file (.pbs)**

For use on shared hardware allocation schemes and outlines the hardware requested, job length and command line arguments and what command line functions to run in order. A typical basic GROMACS PBS file will contain between 5-7 hardware allocation arguments and between 16-18 command line function calls.

## Simulation Workflow:

**Hardware allocation:**

Contained within the Project run script or PBS file the number of required CPU cores, compute nodes, RAM and GPU's are outlined alongside any relevant pre-simulation calls required such as project allocation and loading in any relevant MDS package modules e.g. gromacs/2024.5.

**Molecular Preparation:**

This involves slightly modifying the supplied structure files to ensure compatibility and reliability with the MDS package as well as often translating the file format into that more natively workable with the MDS package. This can include the removal of excess water molecules accidentally captured during x-ray crystallography for lab derived structured or reformatting machine learning generated or mislabelled structures to be compatible with the nomenclature used within the MDS package.

**Defining the Simulation Bounds:**

Here the MDS package is utilised to define simulation constraints and boundaries, this is to ensure and control several factors. First is such that the simulation is maintained to a manageable size for computation, second, that there is not an inflated number of solvent molecules being simulated unnecessarily and lastly so that there is control over the allowed distance between inserted molecules as starting a simulation with two molecules close together will shorten the required simulation time for them to contact and impact with one-another.

**Insertion, Solvation and Charge Balancing:**

Now knowing the bounds of the simulation, the provided molecules can now be inserted within these bounds, whilst ensuring that there are not overlaps between molecules. Additionally, solvent (typically water) is inserted into the empty spaces surrounding these provided structures until no more free space is remaining within the simulation bounds. Finally, the overall system charge is calculated and charged ions (typically chlorine and sodium) are added to offset the system back to neutral charge if desired.

**System Equilibration and Balancing:**

The following three stages comprise of different methods to ensure that the simulation is starting at its lowest possible energy state to prevent any excessive forces from structure and solvent insertion from immediately "blowing up" [29] the simulated system. These methods are typically performed with the inserted structures in a locked immovable state, hence requiring the solvent around these to shift instead and leaving the majority of the structure movement and force calculations for the final simulation stage for capture of the most relevant data.

**Energy Minimisation:**

Involves slightly shifting the position of atoms/molecules that have an excessively high force applied onto them which is often the result of being inserted too closely to other molecules. This is done until all atoms in the system are below a specified energy threshold (typically 500 kJ/mol/nm).

**NVT:**

NVT indicates a simulation variant where the system is held at a constant **N**umber of moles/atoms, **V**olume of the simulation and **T**emperature. This microsimulation is performed with the inserted macromolecules in a fixed state but with the system able to exchange heat with the bounds allowing for an even and fixed temperature throughout the system.

**NPT:**

Very similar to NVT except the simulation is now no longer restrained to a fixed volume and is able to scale the bounding size slightly to achieve a specified and fixed pressure throughout the solvent.

**MD Production/Simulation Run:**
The final and largest stage of MD Simulation is the primary production run where the inserted macromolecules are no longer fixed in place and are permitted to interact with each other and surrounding solvent. This stage, depending on the desired interactions that is being studies, will usually run for magnitudes longer than its prior equilibration stages.

**Analysis:**

Following simulation, a number of analysis methods are available and will require several generated files to complete. A structure file from the beginning of the Production Run alongside a portable trajectory file will be suitable for most analysis methods such as RMSD, RMSF or Radii of Gyration. However, for more detailed analysis that requires specific force and bond information such as H-bond analysis the topology file will also typically be required.

# Overview of MDSGAT GUI System:

MDSGAT can be broken down into three primary modules with each having a varied number of submodules, as with the prior explanation, a summarised series of charts are presented below to encapsulate the stages required to perform MDS using the MDSGAT software with the GROMACS package.

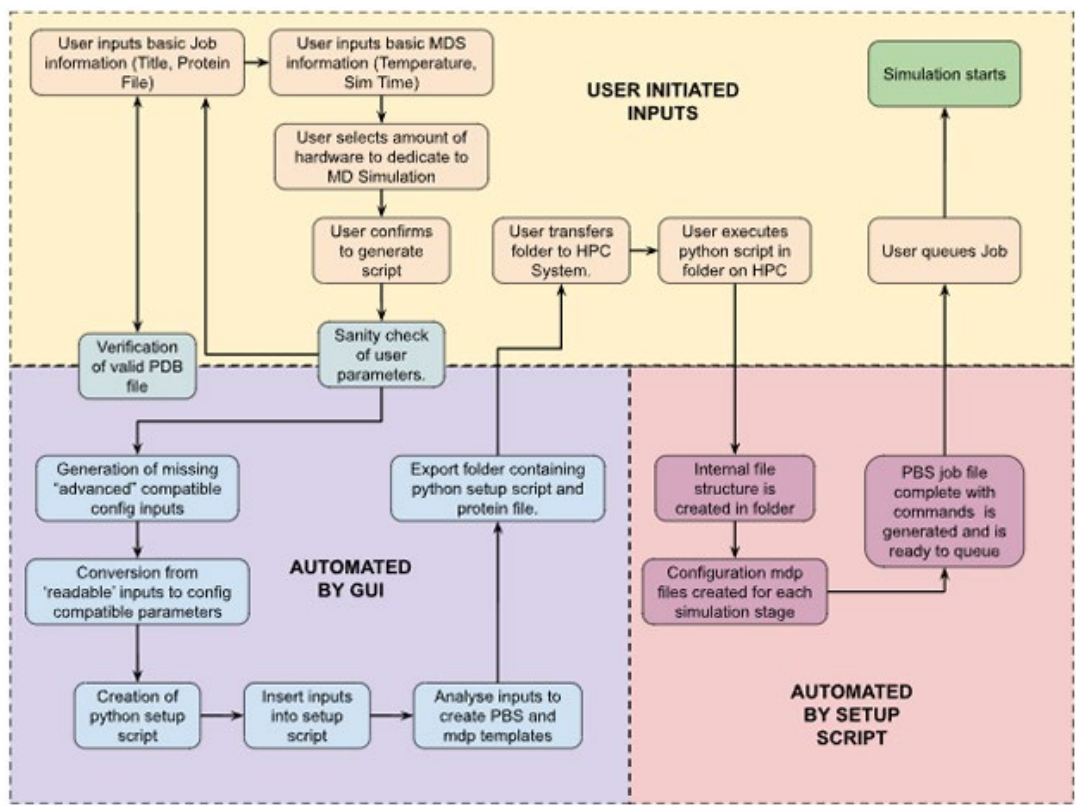

*Figure 2. Summary of the tasks and operations involved when using the MDSGAT software to create and start an MDS simulation.*

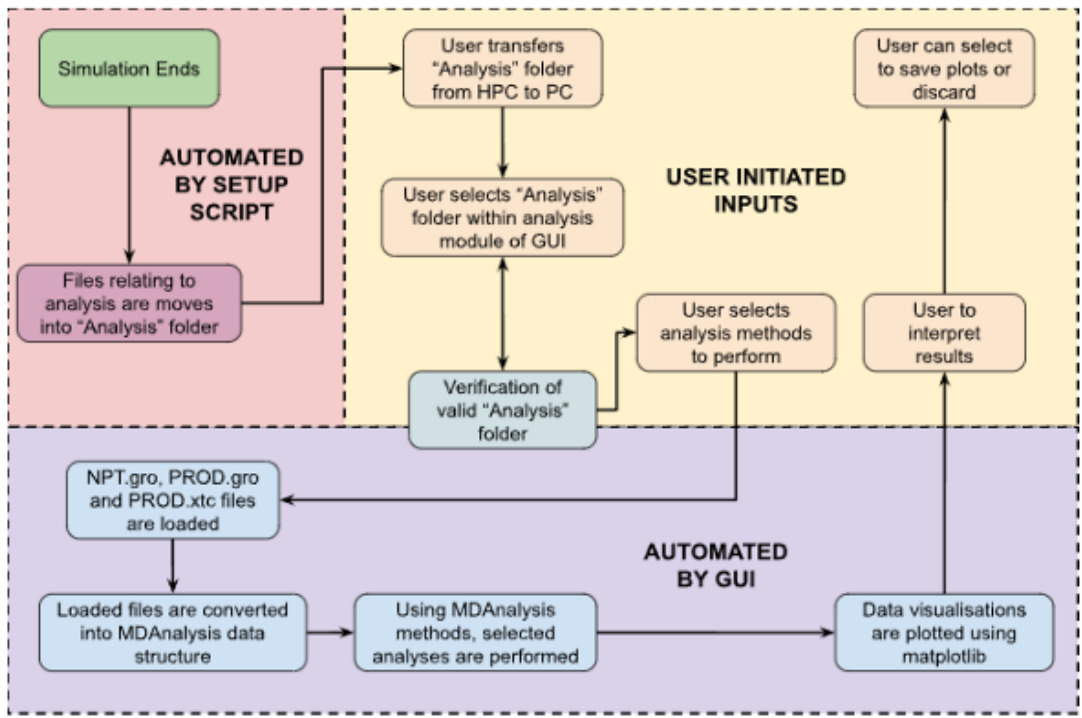

*Figure 3. Summary of the tasks and operations involved when using the MDSGAT software to analyse the results of a successful MDS simulation.*

MDSGAT Core – The core module is responsible for navigation and documentation as well containing the configuration menus for saving computing cluster and project information, whilst also serving as the primary system output for program feedback during tasks.

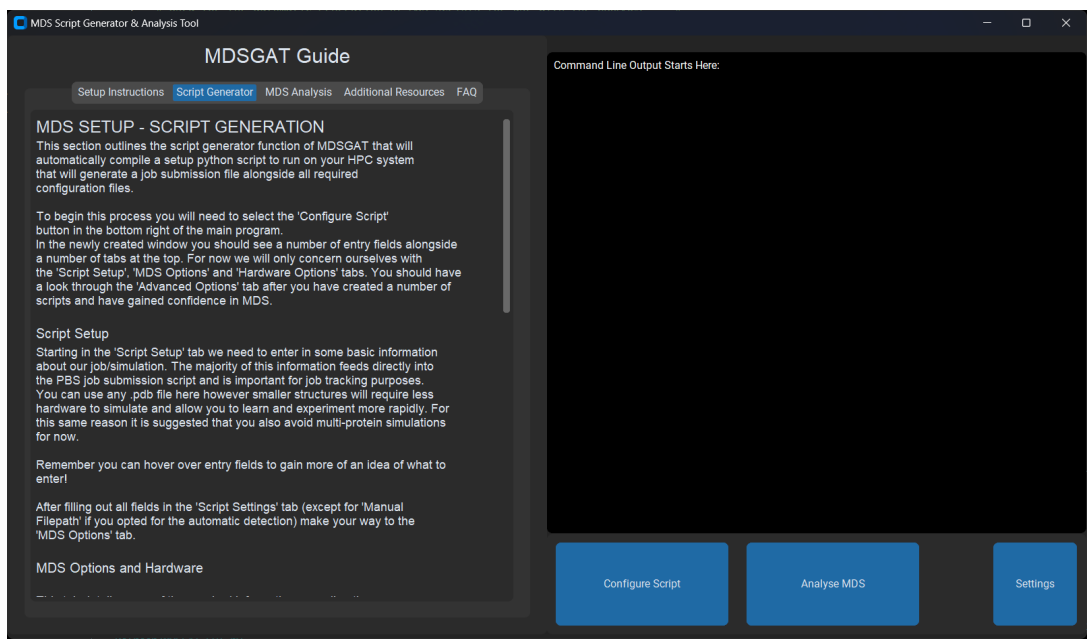

*Figure 4. Initial interface of the core module of MDSGAT depicting the instructional documentation and navigation options.*

MDSGAT Script Generator – The Script Generator module contains three more sub modules; The input submodule is utilised to specify simulation conditions from the user as well as prompt for the desired macromolecule to insert into the simulation. The next submodule is the input converter which will both error check the users input and then convert the human readable inputs into the specific configuration formats required by the configuration files. Finally, the script generator will compile all these inputs and write a standalone formatted python script to generate all of the required files for the simulation once ran and package this into a separate folder with the supplied macromolecule structure file for upload to a computing cluster. This python script is formatted to include all of the relevant simulation conditions at the top of the file for easy inspection and modification should this file be shared.

MDSGAT Analysis – The analysis module contains a number of included analysis methods that are user selectable and displays each of the selected methods on separate a plot grid display after calculation.

## User Input Management:

The MDSGAT Script generator input submodule displays a brief list of input fields to the user along with tooltips and examples of the contents of each, allowing for an informed recommendation to new users, additionally, to obtain the structure data for inserted macromolecules the user is prompted to browse their computer for the location of this file where it is copied into an output folder once the final script has been successfully generated. Many inputs require the user to type a number or text into an entry field such as the job name or runtime temperature whereas for fields with limited options dropdown menus and sliders are utilised. To reduce initial complexity the program only requires users to enter in the most impactful parameters such as simulation temperature, molecule count and simulation runtime and will use a preset list of commonly utilised default values from official documentation [29] and community feedback [30] for the remainder of the configuration unless the user opts into editing these "Advanced Settings"

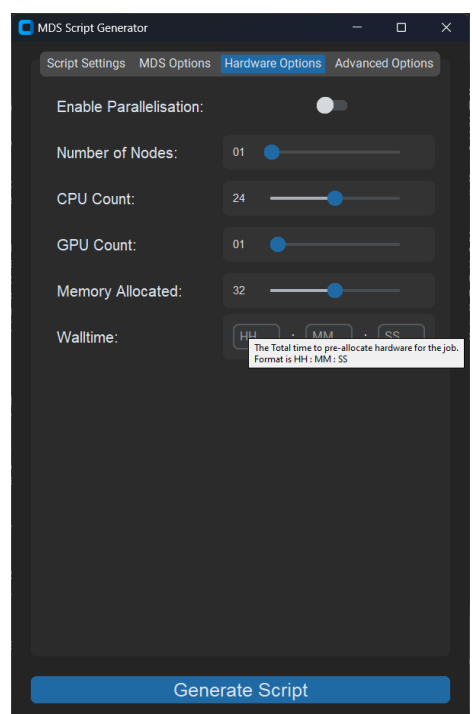

*Figure 5. MDSGAT Interface for part of the user input submodule for script generation with tooltip showing over hovered entry field.*

## Script Generation:

Once the user has finalised their inputs and attempts to generate a setup script the input converter submodule will both check the users' inputs for incorrect formatting or ranges and also convert human readable inputs into suitable configuration values (e.g. None to "-1"). These approved input values are then combined with any autogenerated "Advanced Settings" and passed to the script generator submodule which will then create a new folder containing a python setup file and the originally selected structure files, this newly created python file is designed to be ran on the intended simulating system using python 3 or later. This single 2-file folder is what will be uploaded onto the computing cluster using FTP tools such as FileZilla [31] or shared for collaboration as it contains all the necessary data to replicate the simulation environment. The setup file contains all of the configuration inputs as variables as well as code sections to automatically create the required configuration files (.mdp) and job script files (.pbs) and will format and fill out these files with the corresponding configuration parameters whilst also recognising the folder it is run inside as the simulation runtime directory. Once all the required files have been created then the user can submit their job into the computing queue and wait for the simulation to conclude.

# 5. Post-Simulation Analysis

## Data Retrieval:

Once the simulation has finished running all of the necessary files for analysis will be moved into a single "Analysis" folder in the same directory as the setup script and can be downloaded back to the users local machine using the FTP client of their choice for analysis. Once the data has been downloaded, the Analysis module of MDSGAT can be used to browse for this folder and perform analysis accordingly.

## Analysis Modules:

Currently, MDSGAT supports a modest number of analysis methods including Root Mean Square Deviation (RMSD) [32], Root Mean Square Fluctuation (RMSF) [33], Radii of Gyration (RoG) and Principle Component Analysis (PCA) [34] all supported by the MDAnalysis Python library [35], [36]. RMSD is most commonly utilised for gauging the stability of the simulation and energy shift of the system and is useful for knowing that the simulation results are sound. RMSF is great for analysing which residues within the inserted structure are less stable in the simulated environment and RoG is useful for visualising the rate and total deformation of inserted structures over the duration of the simulation. Each of these analysis methods when selected within the MDSGAT analysis module will be displayed on a matplotlib [37] plot grid and naturally, there are significant benefits to qualitatively viewing the simulation progress using external software such as PyMOL [22], [23], [38] or VMD [39].

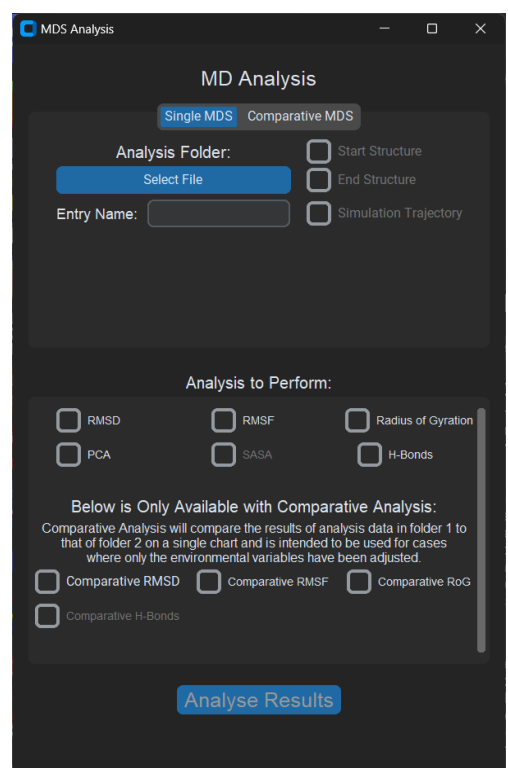

Figure 6. MDSGAT Interface for the user input part of MD Analysis.

# 6. Example Case Study and Workflow

## Installation and Configuration of MDSGAT:

The MDSGAT executable and run files can be located on GitHub [40]. To successfully run MDSGAT simply download and move the MDSGAT directory to your desired location and run the application contained within. This will open the MDSGAT core module containing helpful documentation and navigation buttons. From here on, simply following the included documentation will guide a user towards creating their first MDSGAT simulation script.

# Example – Observing heat stress on Soybean Glycinin G1 (GlyG1) Protein:

## Creating the Setup Script:

Starting with the Protein Databank file for the structure of GlyG1 MDSGAT can be opened, and the Script Generator module can be loaded. Using this menu, information relating to the desired simulation parameters can be filled in. Of note this example will consist of two separate simulations, both will be of a single GlyG1 protein at atmospheric pressure for 10 ns in water with the key differentiator being that the first simulation is ran at 295K and the second being ran at 325K to view the temperature stability of the GlyG1 protein. Once the "Generate Script" button is selected the inputs will be checked and a folder containing the GlyG1 PDB file alongside the python setup script is created.

## Simulation execution on a remote High-Performance Computing (HPC) Cluster:

Using an FTP client the newly created folder can be transferred onto the HPC system and the included script and be executed using python 3. This will generate a number of configuration files and subdirectories alongside the PBS job submission script. Next, to start the simulation all that needs to occur is to submit that job submission script into the job queue and an email will be sent on job start and completion if this feature is enabled on the submitted PBS queue. Once the simulation has been completed all necessary analysis related files will be moved into an "Analysis" folder contained within the same original folder uploaded to the HPC system, then simply using the FTP client again this Analysis folder can be downloaded to the local machine for analysis.

## Analysis and Visualization:

Analysis is performed by opening the MDSGAT program and navigating to the Analysis module. From this module the user can browse for the downloaded "Analysis" folder, select given analysis methods to plot and supply a suitable name for the plot title to display. This stage can be computationally expensive for lower end personal computers and as such can take a number of minutes to complete depending on simulation complexity. Once the analysis is complete a number of plots will be generated and ready for analysis.

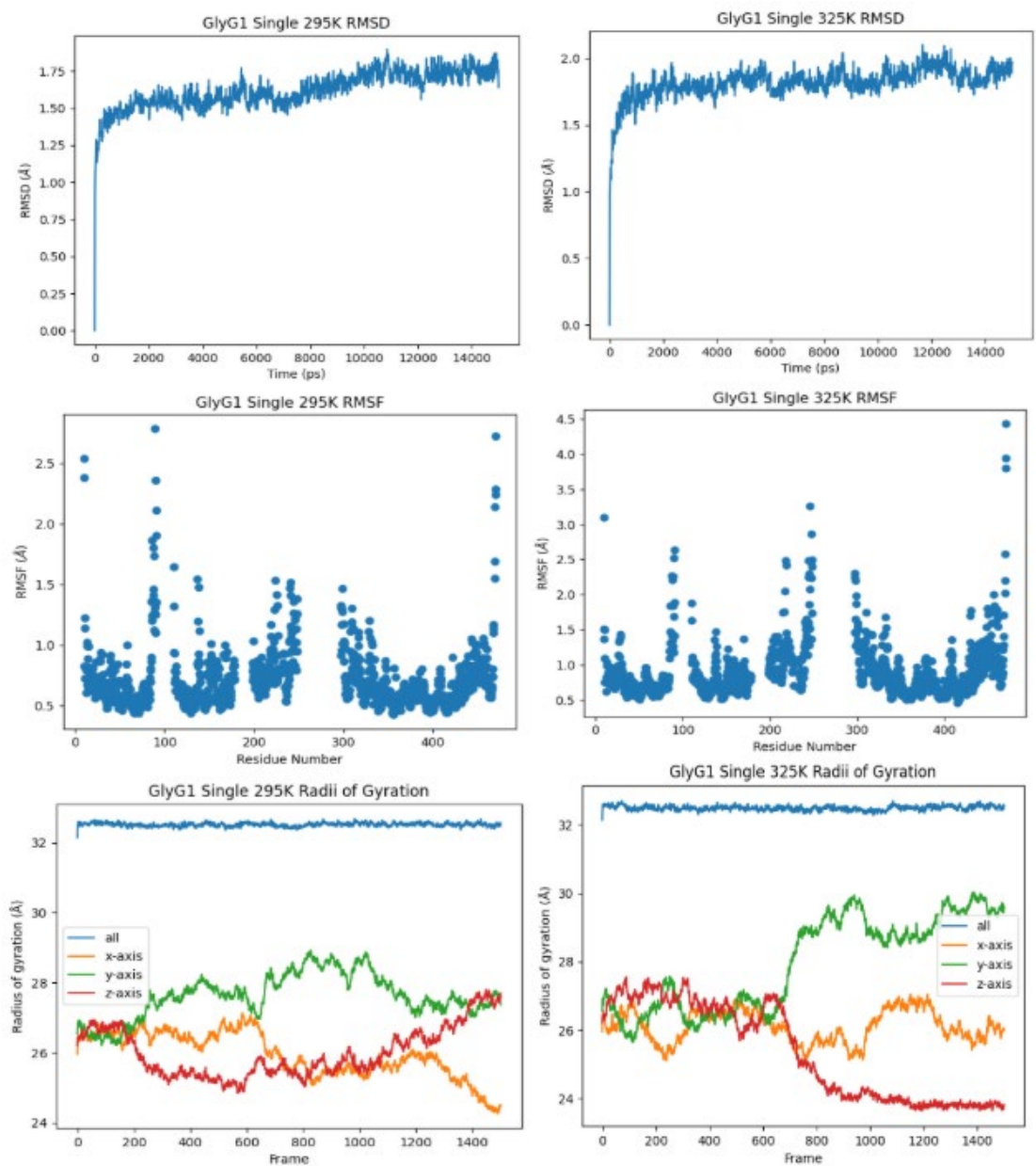

*Figure 7. Example output of the MDSGAT analysis module for the provided GlyG1 environmental example.*

## Rudimentary Findings:

From the generated plots a number of observations are able to be made, and a clear distinction is visible between the 295K and 325K simulations. Firstly, both RMSD plots plateau which is indicative that the simulation has reached an equilibrium state and the GlyG1 protein is relatively stable during this period and hence ready for analysis, additionally the RMSD plots show that the total deviation experienced by GlyG1 is slightly higher in the higher temperature simulation which can be indicative of more motion and is expected. The RMSF plots depict an interesting finding, both plots indicate that the areas around the start and end residues experience much more movement then those in the centre with another spike at the ~100 residue mark, however, the higher 325K simulation shows that instability in the ~250-300 residue region is increased greatly with increased temperatures indicating that this region of the protein is less thermostable than the rest of the residues as it experiences the greatest shift in fluctuation between temperatures. Additionally, the RMSF plot shows that the end of the GlyG1 protein also seems to have its instability impacted more by heat than the starting residue areas. RoG serves as a measure of "compactness" and can typically indicate how well folded a protein remains, looking at the RoG it appears that the lower temperature simulation has a slow and steady unfolding process that seems to plateau fairly early on whereas the higher temperature simulations appears' to depict an event at the ~6 ns mark that leads to a shift towards losing compactness. However, as the total RoG for both simulations remains steady it is safe to assume that the protein retains most of its conformation throughout both simulations.

# 7. Performance Evaluation

## Comparison with Existing Tools:

The MDSGAT suite presents a much simplified and shortened method of creating, running and analysing MD simulations with the vast majority of informative simulations being studied in recent time being capable of setup and running within a few short minutes. The MDSGAT analysis features allows for rapid and simplified analysis of the generated results with meaningful conclusions being easily extrapolated from the presented data. Currently, MDSGAT stands as a viable alternative to the existing market of GUI tools for MDS, presenting similar capabilities with its realised implementation of the GROMACS platform and numerous compatible force fields especially for users seeking a free and open source (FOSS) alternative to the sometimes restrictive and private options. Despite this, MDSGAT does uniquely provide a novel method of preserving all the simulation parameters needed to run and replicate any given simulation generated with the program using a single all-encompassing setup script as a method of reducing the doubt and scepticism some MDS publications are treated with due to concerns of reproducibility and result reliability.

## User Feedback and Experience:

Only a very small number of users have been exposed to the beta variants of MDSGAT including 2 experienced users and 2 new MDS users. The generalised feedback from the new users is that they found the analysis modules incredibly helpful and were even able to successfully submit and run simulations within the same day as starting setup. More experienced users had a number of points on feedback, a common praise was the programs utility to rapidly create simulation 'templates' that they then only had to make minor modifications towards, however, there was a desire for more complex features to be introduced such as additional solvent support, support for multiple species of molecules to be inserted and more refined support for parallel workflows. Additionally, one beta tester reported an over 95% reduction in time spent on simulation setup and refinement for the majority of their simulations ran during the testing period as they we're focussed on performing an array of simple simulations for comparative purposes.

# 8. Conclusion and Future Work

## Summary of Impact:

The aim MDSGAT had during its initial conception is to eventually serve as a single channel for which the majority of MDS needs can be met, and through the creation of such a channel a number of clear benefits are presented to the MD simulation community. Firstly, by consolidating the users and MDS platforms into a single interface community support and aid is dispersed over a smaller footprint better allowing community members to more easily seek assistance from fellow community members. Additionally, a uniform framework and standardised configuration will make MD results more consistent with one another aiding in collaborative research and trust. Furthermore, the implementation and creation of single setup files allows for a user's exact simulation environment to be easily shared and emulated by other community members or peer reviewers, posing as the simplest method of validation for MDS published papers. The final major benefit of a centralised, simplified and open system such as MDSGAT is the accessibility factor for researchers outside of the MD ecosystem, allowing them to easily participate and experiment with MD to further push the boundaries of the potential applications MD is suitable for. The version of MDSGAT displayed in this paper is certainly still some time away from being suitable of achieving all of the aforementioned aims, however, this first release version of MDSGAT is a profound showcase of the direction and intension of realising these objectives.

## Future Enhancements and Road-Mapped Features:

The release of version 1.0 of MDSGAT only constitutes the creation of the foundation for the future of the program to be built on top of with a large number of new features and refinement already showing some progression. Additionally, with the feedback of the beta testing period a number of improvements and expansions are planned for the future of MDSGAT with an estimated feature and version roadmap outlined below:

**Version 2.0** – Non-aqueous solvent support, Integrated FTP support, Expanded analysis capabilities, Codebase improvements, Improved plotting aesthetic and Documentation improvements.

**Version 3.0** – Support for at least one additional MDS package and Project allocation method.

**Version 4.0** – Support for native protein rendering, trajectory visualisation, more advanced control over initial starting conditions.

**Version 5.0** – Simulation checkpointing and variable environment simulation.